\documentclass[aps,pre,preprint,amsmath,amssymb]{revtex4}

\usepackage{graphicx}
\usepackage{dcolumn}
\usepackage{bm}

\usepackage{amsmath}
\usepackage{placeins}

\newcommand{\mbi}[1]{\mbox{\boldmath $#1$}}
\newcommand{\sgn}{\mbox{sgn}}

\begin{document}
\preprint{}
\title{Statistical mechanics of lossy compression for non-monotonic multilayer perceptrons}
\author{Florent Cousseau}
\affiliation{
Graduate School of Frontier Sciences, University of Tokyo, Chiba 277-5861, Japan}
\email{florent@mns.k.u-tokyo.ac.jp}
\author{Kazushi Mimura}
\affiliation{
Faculty of Information Sciences, Hiroshima City University, Hiroshima 731-3194, Japan}
\email{mimura@hiroshima-cu.ac.jp}
\author{Toshiaki Omori}
\affiliation{Brain Science Institute, RIKEN, Saitama 351-0198, Japan}
\author{Masato Okada}
\affiliation{
Graduate School of Frontier Sciences, University of Tokyo, Chiba 277-5861, Japan \\
Brain Science Institute, RIKEN, Saitama 351-0198, Japan
}
\date{\today}
\begin{abstract}
\label{abst}
A lossy data compression scheme for uniformly biased Boolean messages
is investigated via statistical mechanics techniques. We utilize tree-like
committee machine (committee tree) and tree-like parity machine (parity tree)
whose transfer functions are non-monotonic. The scheme performance at the
infinite code length limit is analyzed using the replica method. Both
committee and parity treelike networks are shown to saturate the Shannon
bound. The AT stability of the Replica Symmetric solution is analyzed,
and the tuning of the non-monotonic transfer function is also discussed.
\end{abstract}

\maketitle

\section{introduction}

The tools of statistical mechanics have been successfully applied in
several problems of information theory in recent years. In
particular, in the field of error correcting codes
\cite{Sourlas1989,Kabashima2000,Nishimori1999,Montanari2000},
spreading codes \cite{Tanaka2001,Tanaka2005},
and compression codes \cite {Murayama2003,Murayama2004,Hosaka2002,Hosaka2005,Mimura2006},
statistical mechanical techniques have shown great potential. The
present paper uses similar techniques to investigate a lossy
compression scheme. Lossless compression, which was first pointed
out by the pioneering paper of Shannon \cite{Shannon1948}, has been
widely studied for many years. After much effort, a set of very good
codes have been designed and practical implementations have been
proposed \cite{Gallager1962,MacKay1997,Richardson2001}.
Lossy compression, on the other hand, was also first studied by
another paper of Shannon \cite{Shannon1959}. A lot of practical lossy compression schemes were
developed over the years (for example JPEG compression, MPEG compression etc.) but at the
present time, none of these schemes saturate the Shannon bound given
by the \textit{rate-distortion theorem}.
Nevertheless, several theoretical schemes which reach this optimal bound have already been proposed. Recently, Shannon optimal codes based on sparse systems have been discovered
\cite{Murayama2003,Matsunaga2003,Martinian2006,Wainwright2007} and it is now the general tendency to use such kinds of systems. These codes saturate
the Shannon bound asymptotically, (i.e.: for an infinite codeword length), and in the dense generating matrix limit (but low connectivity sparse matrix already gives near Shannon performance).
However, there is still a lot of
work to be done for densely connected systems.
One of such systems
is given by using perceptron-based decoder.
There have been some
recent studies on the encoding problem of such schemes using the
belief propagation (BP) algorithm and the results seems promising
\cite{Hosaka2006}. The foundations of this encoding method for such lossy compression
schemes was originally put forward by Murayama using the TAP
equations applied to Sourlas-type codes \cite{Murayama2004}.
It is important to study a wide class of decoder
to extract a pool of schemes which can give near Shannon bound
performance prior to fully investigate the encoding problem. The
study of such schemes could gives interesting clues on how the lossy
compression process works, and it might also help to pinpoint
some essential features a scheme should possess in order
to achieve Shannon optimal performance.

This paper extends the framework
introduced in \cite{Hosaka2002,Hosaka2005,Mimura2006} and studies three
different decoders based on a non-monotonic multilayer perceptron.
Hosaka et al. studied the simple perceptron network featuring a non-monotonic transfer function
in order to have a mirror symmetry property in their model
(i.e.: $f(u)=f(-u)$). This was motivated by the belief that the Edwards-Anderson order
parameter should be zero to reach the Shannon bound. Consequently, if one codeword
$\mbi{s}$ is optimal (note that here optimal denotes a codeword which gives the minimum achievable distortion for the concerned scheme), $-\mbi{s}$ is also optimal. Then, they
show that for an infinite length codeword, their scheme effectively saturates the Shannon bound.
Next, one interesting feature of the model proposed by Mimura et al. \cite{Mimura2006} is to
increase the number of optimal codewords by using a multilayer decoder network. The number
of optimal codeword is function of the number of hidden units $K$ in the decoder network (for example, in
their parity-tree model with an even number of hidden units, there are at least $2^{K-1}$ optimal codewords).
Thus, one can expect that finding an optimal codeword becomes more and more easy as the number of hidden units increases. Nevertheless, their model deal only with unbiased messages.
The main advantage of the model proposed in this paper is to combine the benefits of Hosaka et al. model (mirror symmetry and ability to deal with biased messages) with the benefits of Mimura et al. model (increasing number of optimal codewords with the number of hidden units).
By studying three different schemes we would like to extract some essential characteristics a good lossy compression framework should possess.
Finally, the Almeida-Thouless (AT) stability of the obtained solutions is also studied and
presents very good properties with almost no unstable part.

The paper is organized as follows. Section II introduces the framework of
lossy compression. Section III exposes our model. Section IV presents the mathematical
tools used to evaluate the performance of the present scheme. Section V states some results
concerning the validity of the obtained solutions and section VI is devoted to conclusion
and discussion.

\section{Lossy Compression}

\begin{figure}[h]
  \vspace{0mm}
  \begin{center}
  \includegraphics[width=0.85\linewidth,keepaspectratio]{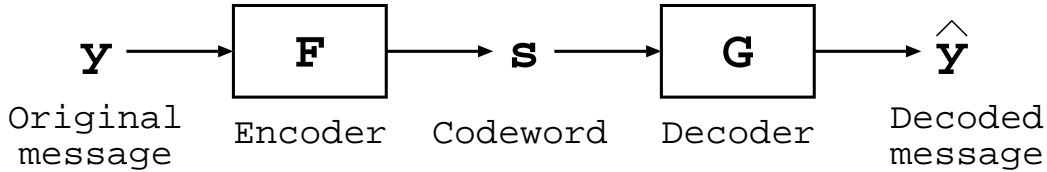}
  \end{center}
  \caption{Rate distortion encoder and decoder.}
  \label{fig:framework}
  \vspace{0mm}
\end{figure}

Let us begin by introducing the framework of lossy data compression \cite{Cover1991}.
Let $\mbi{y}$ be a discrete random variable defined on a source alphabet $\mathcal{Y}$.
An original source message is composed of $M$ random variables,
$\mbi{y} = (y^1,\ldots,y^M) \in \mathcal{Y}^M$, and compressed into a shorter expression.
The encoder compresses the original message $\mbi{y}$ into a codeword $\mbi{s}$,
using the transformation $\mbi{s}=\mathcal{F}(\mbi{y}) \in \mathcal{S}^N$, where $N<M$.
The decoder maps this codeword $\mbi{s}$ onto the decoded message $\hat{\mbi{y}}$,
using the transformation $\hat{\mbi{y}}=\mathcal{G}(\mbi{s}) \in \hat{\mathcal{Y}}^M$. The
encoding/decoding scheme can be represented as in Figure \ref{fig:framework}.
In this case, the code rate is defined by $R=N/M$.
A distortion function $d$ is defined as a mapping
$d:\mathcal{Y} \times \hat{\mathcal{Y}}\rightarrow \mathbb{R}^+$.
For each possible pair of $(y,\hat{y})$, it associates a positive real number.
In most of the cases, the reproduction alphabet $\hat{\mathcal{Y}}$
is the same as the alphabet $\mathcal{Y}$ on which the original message $\mbi{y}$
is defined.

Hereafter, we set $\hat{\mathcal{Y}}=\mathcal{Y}$, and we use the Hamming
distortion as the distortion function of the scheme. This distortion function is given by
\begin{equation}
d(y,\hat{y})=
\left\{
          \begin{array}{ll}
            0, & \qquad y=\hat{y}, \\
            1, & \qquad y \neq \hat{y}, \\
          \end{array}
        \right.
\end{equation}
so that the quantity $d(\mbi{y},\hat{\mbi{y}}) = \sum_{\mu=1}^M d(y^{\mu},\hat{y^{\mu}})$
measures how far the decoded message $\hat{\mbi{y}}$ is from
the original message $\mbi{y}$.
In other words, it records the error made on the original message
during the encoding/decoding process. The probability of error distortion
can be written $E[d(y,\hat{y})]=P[y \neq \hat{y}]$ where $E$ represents
the expectation. Therefore, the distortion associated with the code is defined
as $D=E[\frac 1 M d(\mbi{y},\hat{\mbi{y}})]$, where the expectation is taken with respect to
the probability distribution $P[\mbi{y},\hat{\mbi{y}}]$. $D$ corresponds to the
average error per variable $\hat{y}^{\mu}$. Now we defined a rate
 distortion pair $(R,D)$ and we said that this pair is achievable if there exist a
 coding/decoding scheme such that when $M \to \infty$ and $N \to \infty$
 (note that the rate $R$ is kept finite), we have
$E[\frac 1 M d(\mbi{y},\hat{\mbi{y}})] \leq D$. In other words, a rate distortion pair
$(R,D)$ is said to be achievable if there exist a pair $(\mathcal{F},\mathcal{G})$ such that
$E[\frac 1 M d(\mbi{y},\hat{\mbi{y}})] \leq D$ in the limit
$M \to \infty$ and $N \to \infty$.

The optimal compression performance that can be obtained
in the framework of lossy compression is given by the so-called
\textit{rate distortion function} $R(D)$ which gives the
best achievable code rate $R$ as a function of $D$ \cite{Cover1991}.
However, despite the fact that the best achievable performance is known,
no clues are given about how to construct such an optimal compression scheme.
Moreover, finding explicitly the expression of the rate distortion function
is, in general cases, not possible.

Nonetheless, for the special case of uniformly biased Boolean messages in which
each component is generated independently by the same probability distribution
$P[y=0]=1-P[y=1]=p$, it is possible to calculate analytically the rate distortion
function $R(D)$, which becomes
\begin{equation}
R(D) = H_2 (p) - H_2 (D), \label{rate-distortion}
\end{equation}
where $H_2 (x) = -x \log_2 x - (1-x) \log_2 (1-x)$. In the sequel, we restrict ourselves
to this particular case (i.e. : $P[y=0]=1-P[y=1]=p$ and $\mathcal{Y}=\{0,1\}$).

\section{Compression using non-monotonic multilayer perceptrons}

In this section we introduce our compression scheme. To make the calculations
compatible with the statistical mechanics framework, let us map the Boolean representation
$\{0,1\}$ to the Ising representation $\{-1,1\}$ by means of the mapping
$\sigma = (-1)^{\rho}$, where $\sigma$ is the Ising variable and $\rho$ is the
Boolean one.
On top of that, we set $\mathcal{Y}=\mathcal{S}=\mathcal{\hat{Y}}=\{-1,1\}$.
Since we consider that all the $y^{\mu}$ are generated independently
by an identical biased binary source, we can easily write the corresponding
probability distribution,
\begin{equation}
P[y^{\mu}] = p \delta (1-y^{\mu}) + (1-p) \delta (1+y^{\mu}). \label{ydistrib}
\end{equation}
Next we define the decoder of the compression scheme. We use a non-linear
transformation $\mathcal{G} : \mathcal{S}^N \to \mathcal{\hat{Y}}^M$ which
associates a codeword $\mbi{s} \in \mathcal{S}^N$ with a sequence
$\hat{\mbi{y}} \in \hat{\mathcal{Y}}^M$.
For a given original message $\mbi{y}$, the encoder is simply defined as follows,
\begin{equation}
\mathcal{F}(\mbi{y}) \equiv \underset{\mbi{\hat{s}}}{\text{argmin}} \,
d(\mbi{y},\mathcal{G}(\mbi{\hat{s}})). \label{encoder}
\end{equation}

For the non-linear transformation $\mathcal{G}$, we utilize non-monotonic multilayer
perceptrons. The codeword $\mbi{s}$ is split down into $N / K$-dimensional $K$ disjoint
vectors $\mbi{s}_1, \ldots , \mbi{s}_K \in \mathcal{S}^{N/K}$ so that $\mbi{s}$ can be written
$\mbi{s}=(\mbi{s}_1, \ldots , \mbi{s}_K)$. In this paper, we will focus on three different architectures for the non-monotonic multilayer perceptrons. There are the followings :

(I) Multilayer parity tree with non-monotonic hidden units. Its output is written
\begin{equation}
\hat{y}^{\mu} (\mbi{s}) \equiv
\prod_{l=1}^K f_k \left( \sqrt{\frac {K} {N}} \, \mbi{s}_l \cdot \mbi{x}_l^{\mu} \right) .
\label{parity}
\end{equation}

(II) Multilayer committee tree with non-monotonic hidden units. Its output is written
\begin{equation}
\hat{y}^{\mu} (\mbi{s}) \equiv
\sgn \left( \sum_{l=1}^K f_k \left[ \sqrt{\frac {K} {N}} \, \mbi{s}_l \cdot \mbi{x}_l^{\mu} \right] \right) .
\label{committee1}
\end{equation}
Note that in this case, if the number of hidden units $K$ is even,
then there is a possibility to get $0$ for the argument of the sign
function. We avoid this uncertainty by considering only an odd
number of hidden units for the committee tree with non-monotonic
hidden units in the sequel.

\begin{figure}[h]
  \vspace{0mm}
  \begin{center}
  \includegraphics[width=0.6\linewidth,keepaspectratio]{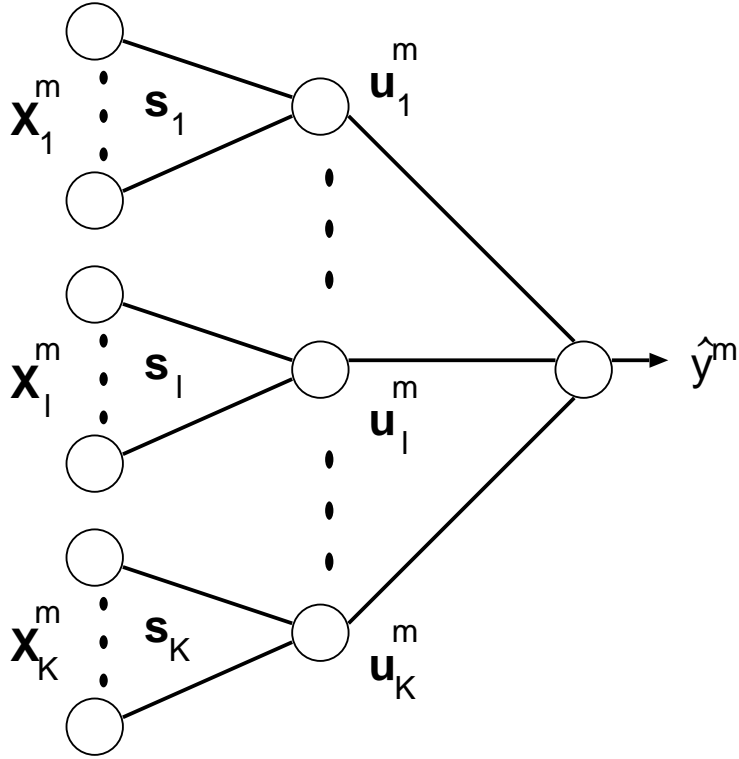}
  \end{center}
  \caption{General architecture of the treelike multilayer perceptron with $N$ input units
  and $K$ hidden units.}
  \label{fig:net}
  \vspace{0mm}
\end{figure}

(III) Multilayer committee tree with a non-monotonic output unit. Its output is written
\begin{equation}
\hat{y}^{\mu} (\mbi{s}) \equiv
f_k \left( \sqrt {\frac 1 K}\sum_{l=1}^K
\sgn \left[ \sqrt{\frac {K} {N}} \, \mbi{s}_l \cdot \mbi{x}_l^{\mu}
\right] \right) .
\label{committee2}
\end{equation}
In each of these structure, $f_k$ is a non-monotonic function of a real parameter $k$ of the
form
\begin{equation}
f_k (x) =
\begin{cases}
1 \quad \text{if} \quad |x| \leq k
\\
-1 \quad \text{if} \quad |x| > k
\end{cases}
\end{equation}
and the vectors $\mbi{x}^{\mu}_l$ are fixed $N/K$-dimensional independent vectors
uniformly distributed on $\{-1,1\}$. The $\sgn$ function denotes the sign function taking $1$
for $x \geq 0$ and $-1$ for $x <0$. Each of this architecture applies a
different transformation to the codeword $\mbi{s}$.
The general architecture of these perceptrons based decoders is shown in Figure \ref{fig:net}.
Note that we can also consider a decoder based on a committee-tree where both the hidden-units
and the output unit are non-monotonic. However, this introduces an extra-parameter (we will
have one threshold parameter for the hidden-units, and one for the output unit) to tune
and the performance should not change drastically. For simplicity, we restrict our study
to the above three cases only.

Now let us introduce $\mathcal{H}$, an energy function of the system,
\begin{equation}
\mathcal{H} (\mbi{y},\mbi{\hat{y}}(\mbi{s})) = d(\mbi{y},\mbi{\hat{y}} (\mbi{s})).
\end{equation}
This energy function $\mathcal{H}$ is clearly minimized for a codeword $\mbi{s}$ which
satisfies equation (\ref{encoder}). Furthermore, in the Ising representation, the Hamming
distance $d$ takes a simple form
\begin{equation}
d(x,y) = 1 - \Theta (xy),
\end{equation}
where $\Theta$ denotes the unit step function which takes $1$ for $x \geq 0$ and 0 for
$x<0$.

The encoding phase can be viewed as a classical perceptron learning problem, where one tries
to find the weight vector $\mbi{s}$ which minimizes the energy function $\mathcal{H}$ for
the original message $\mbi{y}$ and the random input vector $\mbi{x}$.
The vector $\mbi{s}$ which achieve this minimum
gives us the codeword to be send to the decoder. Therefore, in the case
of a lossless compression scheme
(i.e.: $D=0$), evaluating the rate distortion property of the present scheme is equivalent
to finding the number of couplings $\mbi{s}$ which satisfies the input/output relation
$\mbi{x}^{\mu} \mapsto y^{\mu}$. In other words, this is equivalent to the
calculation of the storage capacity of the network \cite{Gardner1988,Krauth1989}.

The choice of parity-tree based and committee-tree based network is motivated by the thorough literature available on this kind of networks. Parity and committee machines have been intensively studied (see \cite{Engel2001} for an overview) by the machine-learning community over the years.
The techniques used to calculate the storage capacity of such networks gives us a starting point for our analytical evaluation of the typical performance of the above schemes.

\section{Analytical Evaluation}

We analyze the performance of these three different compression schemes using the tools of
statistical mechanics. We first define the following partition function,
\begin{equation}
Z(\beta,\mbi{y},\mbi{x}) = \sum_{\mbi{s}}
\exp \left[ -\beta \mathcal{H} (\mbi{y},\mbi{\hat{y}}(\mbi{s})) \right] ,
\end{equation}
where the sum over $\mbi{s}$ represents the sum over all the possible states for the vector
$\mbi{s}$. $\beta$ denotes the inverse temperature parameter. Such a partition function
can be identified with the partition function of a spin glass system with dynamical
variables $\mbi{s}$ and quenched variables $\mbi{x}$. For a fixed Hamming distortion
$MD = E [ d(\mbi{y},\mbi{\hat{y}}) ]$, the average of this partition
function over $\mbi{y}$ and $\mbi{x}$ naturally contains all the interesting typical
properties of the scheme such as the entropy. However, evaluating this average is
hard and we need some technique to investigate it. In this paper we use the so-called
\textit{Replica Method} in order to calculate the average of the partition function.
In the case of such a discrete system,
the entropy should not be negative so that the zero entropy criterion
(see \cite{Krauth1989}) gives us the best achievable code rate limit.
The replica method's calculations to obtain
the average of the partition function
$\left< Z(\beta,\mbi{y},\mbi{x}) \right>_{\mbi{y},\mbi{x}}$ are detailed in Appendix
\ref{appendix.ReplicaMethod}.

\subsection{Replica symmetric solution for the parity tree with non-monotonic hidden units}

In the lossy compression scheme using parity tree with non-monotonic hidden units (\ref{parity}),
the replica symmetric free energy is given by
\begin{eqnarray}
f(\beta,R,k) & = & - \frac 1 {\beta} \Big( p \ln \left[ e^{-\beta}
+ (1 - e^{-\beta}) A_k \right]
\nonumber \\
&& + (1-p) \ln \left[ e^{-\beta}
+ (1 - e^{-\beta}) (1-A_k) \right]
\nonumber \\
&& + R \ln 2 \Big), \label{parityRSenergy}
\end{eqnarray}
where
\begin{eqnarray}
A_k & = & \frac 1 2 + \frac 1 2 \left[ 1 - 4 H(k) \right]^K,
\nonumber \\
H(k) & = & \int_{k}^{+ \infty} \frac {e^{- t^2 / 2}} {\sqrt{2 \pi}} dt.
\end{eqnarray}
The internal free energy is
\begin{eqnarray}
u(\beta,k) & = & p \frac {e^{-\beta} ( 1 -  A_k)} {e^{-\beta} + (1 -e^{-\beta}) A_k}
\nonumber \\
&& + (1-p) \frac { e^{-\beta} A_k} {e^{-\beta} + (1 -e^{-\beta}) (1-A_k)}.
\end{eqnarray}
Minimizing the free energy with respect to $A_k$, taking the zero
temperature limit $\beta \to \infty$ and using the identity
(\ref{Uident}) gives
\begin{eqnarray}
A_k & = & \frac {p-D} {1-2D} \label{Akparity}
\\
D & = & \frac {e^{-\beta}} {1+e^{-\beta}}. \label{betaparity}
\end{eqnarray}
Finally, using the zero entropy criterion, one can get
\begin{equation}
R = H_2 (p) - H_2 (D),
\end{equation}
which is identical to the rate-distortion function (\ref{rate-distortion}).

However, this minimum is reached under the conditions given by equations $(\ref{Akparity})$
and $(\ref{betaparity})$. Since $D$ is fixed, the condition given by the
equation $(\ref{betaparity})$ is easily
satisfied by choosing the proper
inverse temperature parameter $\beta = \ln [(1-D)/D]$.
On the other hand, the condition given
by equation $(\ref{Akparity})$ is satisfied by properly tuning the parameter $k$ of the
non-monotonic function $f_k$. Let us denote the optimal $k$ which satisfies
equation $(\ref{Akparity})$ by $\hat{k}$.
In the case of the parity tree, this optimal $\hat{k}$ is
such that the following equation becomes true
\begin{equation}
H(\hat{k}) = \frac 1 4 \left( 1 - \sqrt[K]{\frac {2p-1} {1-2D}} \right) . \label{khatparity}
\end{equation}
In this paper we consider that
$(p,D) \in \{ [ 0, \frac 1 2 ] \}^2$, therefore, one can
easily show that for $p \neq \frac 1 2$ then $\frac {2p-1} {1-2D}$
is negative which implies that
there is no real solution for the above equation if we have
 an even number of hidden units $K$ (because of the $K$-th root).
However, we can also
consider the case where $(p,D) \in \{ [\frac 1 2 , 1 ] \times [ 0, \frac 1 2 ] \}$
without any change (the probability of $y=1$ and $y=-1$ are just inverted)
and in this case  $\frac {2p-1} {1-2D}$ is positive which implies that
there is always a solution for an any value of $K$. The above problem is just a consequence
of the definition of $p$, but is not related to the model.
So in the case of the parity tree,
$\hat{k}$ always exists independently of the value of $K$.

Finally, since we used the replica symmetric (RS) ansatz, we have to verify
the Almeida-Thouless (AT) stability of the solution to confirm its validity. This is done
in the next section.

\subsection{Replica symmetric solution for the committee tree with non-monotonic hidden units}

In the lossy compression scheme using committee tree with non-monotonic hidden units
(\ref{committee1}), the replica symmetric free energy is given by
\begin{eqnarray}
f(\beta,R,k) & = & - \frac 1 {\beta} \Big( p \ln \left[ e^{-\beta}
+ (1 - e^{-\beta}) B_k \right]
\nonumber \\
&& + (1-p) \ln \left[ e^{-\beta}
+ (1 - e^{-\beta}) (1-B_k) \right]
\nonumber \\
&& + R \ln 2 \Big), \label{committeeHRSenergy}
\end{eqnarray}
where
\begin{eqnarray}
B_k & = & \sum_{\tau_l = \pm 1} \left\{ \Theta \left[ \sum_{l=1}^K \tau_l \right]
\prod_{l=1}^K \left[ \frac {1 + \tau_l (1 - 4 H [ k ])} {2} \right] \right\}.
\end{eqnarray}
The sum other $\tau_l$ represent the sum over each possible state for the dummy variable
$\tau_l$ ($\tau_l = \pm 1$).
The internal free energy is
\begin{eqnarray}
u(\beta,k) & = & p \frac {e^{-\beta} ( 1 - B_k)} {e^{-\beta} + (1 -e^{-\beta}) B_k}
\nonumber \\
&& + (1-p) \frac {e^{-\beta} B_k} {e^{-\beta} + (1 -e^{-\beta}) (1-B_k)}.
\end{eqnarray}
As in the parity tree case, after minimizing the free energy with respect to $B_k$, taking the
zero temperature limit $\beta \to \infty$ and using the identity
(\ref{Uident}), we obtain
\begin{eqnarray}
B_k & = & \frac {p-D} {1-2D} \label{Akcommitteeoutput}
\\
D & = & \frac {e^{-\beta}} {1+e^{-\beta}}. \label{betacommitteehidden}
\end{eqnarray}
Finally, using the zero entropy criterion, one can get
\begin{equation}
R = H_2 (p) - H_2 (D),
\end{equation}
which is identical to the rate-distortion function (\ref{rate-distortion}).
However, here it is not easy to discuss the existence of an optimal $\hat{k}$
which satisfies the condition given by equation $(\ref{Akcommitteeoutput})$.
Such an optimal $\hat{k}$ satisfies the following equation
\begin{equation}
\sum_{\tau_l = \pm 1} \left\{ \Theta \left[ \sum_{l=1}^K \tau_l \right]
\prod_{l=1}^K \left[ \frac {1 + \tau_l (1 - 4 H [ \hat{k} ])} {2} \right] \right\} = \frac {p-D} {1-2D}. \label{khatcommitteehidden}
\end{equation}
We will discuss a little bit more on this existence problem in the next section, when
we check the AT stability of the RS solution.

\subsection{Replica symmetric solution for the committee tree with a non-monotonic output unit}

In the lossy compression scheme using committee tree with a non-monotonic output unit
(\ref{committee2}), the replica symmetric free energy is given by
\begin{eqnarray}
f(\beta,R,k) & = & - \frac 1 {\beta} \Big( p \ln \left[ e^{-\beta}
+ (1 - e^{-\beta}) C_k \right]
\nonumber \\
&& + (1-p) \ln \left[ e^{-\beta}
+ (1 - e^{-\beta}) (1-C_k) \right]
\nonumber \\
&& + R \ln 2 \Big), \label{committeeORSenergy}
\end{eqnarray}
where
\begin{eqnarray}
C_k & = & 2^{-K} \sum_{l = 0}^K \binom{K}{l} \Theta \left[ k^2 - \frac 1 K
(2l-K)^2 \right] .
\end{eqnarray}
The term $\binom{n}{l}$ denotes the binomial coefficient.
The internal free energy is
\begin{eqnarray}
u(\beta,k) & = & p \frac {e^{-\beta} (1 - C_k)} {e^{-\beta} + (1 -e^{-\beta}) C_k}
\nonumber \\
&& + (1-p) \frac {e^{-\beta} C_k} {e^{-\beta} + (1 -e^{-\beta}) (1-C_k)}.
\end{eqnarray}
As in the parity tree case, after minimizing the free energy with respect to $C_k$, taking the
zero temperature limit $\beta \to \infty$ and using the identity
(\ref{Uident}), we obtain
\begin{eqnarray}
C_k & = & \frac {p-D} {1-2D}
\\
D & = & \frac {e^{-\beta}} {1+e^{-\beta}}. \label{betacommitteeoutput}
\end{eqnarray}
Finally, using the zero entropy criterion, one can get
\begin{equation}
R = H_2 (p) - H_2 (D),
\end{equation}
which is identical to the rate-distortion function (\ref{rate-distortion}).
However, here also, it is not easy to discuss the existence of an optimal $\hat{k}$.
Such an optimal $\hat{k}$ satisfies the following equation
\begin{equation}
2^{-K} \sum_{l = 0}^K \binom{K}{l} \Theta \left[ \hat{k}^2 - \frac 1 K
(2l-K)^2 \right] = \frac {p-D} {1-2D}. \label{khatcommitteeoutput}
\end{equation}
This existence problem is discussed later,
when checking the AT stability of the RS solution.

\section{Almeida-Thouless stability of the replica symmetric solution}

In this section we check the AT stability (see \cite{Almeida1978}) of the RS solution
of each scheme. We use the same method as in \cite{Gardner1988,Mimura2006}. The main
mathematical points of the AT stability study are given in Appendix \ref{appendix.AT}.

\subsection{AT stability for the parity tree with non-monotonic hidden units}

In the case of a parity tree with non-monotonic hidden units, we find
\begin{eqnarray}
P & = & \frac 8 {\pi} R^{-1} k^2 e^{- k^2} (e^{\beta} - 1)^2
\nonumber \\
&& \times \left< \left[ \frac {[1 -4H(k)]^{K-1}}
{(e^{\beta} + 1) + (e^{\beta} - 1) y [1 -4H(k)]^K} \right]^2 \right>_y,
\nonumber \\
\nonumber \\
Q &= & R=P^{\prime} = Q^{\prime} = R^{\prime}=0.
\end{eqnarray}
Therefore, using equation (\ref{ATcrit}), the RS stability criterion is given by
\begin{eqnarray}
R & > & \frac 8 {\pi} K \hat{k}^2 e^{- \hat{k}^2} (e^{\beta} - 1)^2
\nonumber \\
&& \times \left< \left[ \frac {[1 -4H(\hat{k})]^{K-1}} {(e^{\beta} + 1) + (e^{\beta} - 1) y
[1 -4H(\hat{k})]^K} \right]^2 \right>_y,
\end{eqnarray}
where $\beta$ is given by (\ref{betaparity}), and where $\hat{k}$ satisfies equation (\ref{khatparity}). $<\ldots>_y$ denotes the expectation with respect to (\ref{ydistrib}).

For $p=\frac 1 2$, that is to say for an unbiased message $\mbi{y}$, $\hat{k}$ satisfies
the equation $H(\hat{k})=\frac 1 4$ which implies $[1-4H(\hat{k})]=0$ and so
the AT line is given by the line $R=0$. Consequently, for unbiased
messages, the RS solution is always AT stable.

\begin{figure}[h]
  \vspace{0mm}
  \begin{center}
  \includegraphics[width=0.8\linewidth,keepaspectratio]{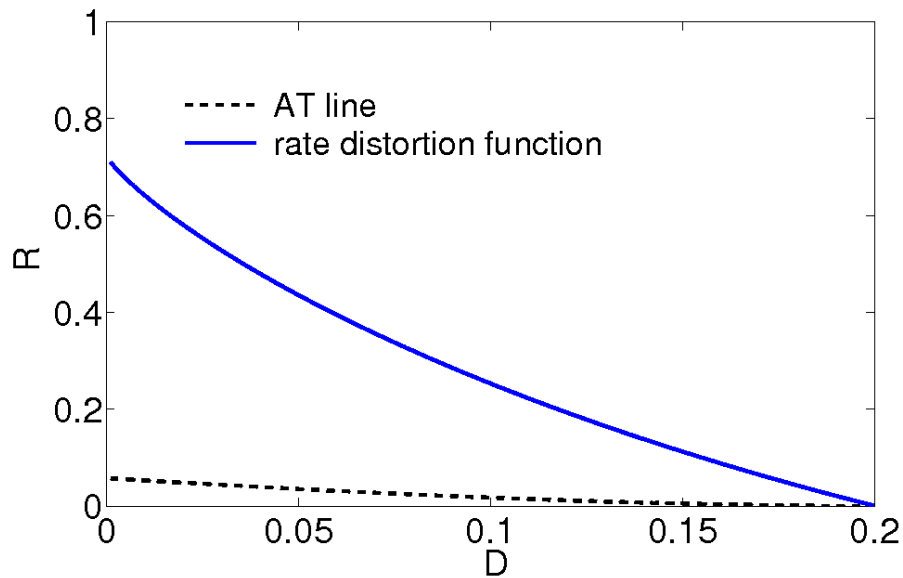}
  \end{center}
  \caption{AT line and rate distortion function for the parity tree with $3$ hidden units.
  The rate distortion performance of the parity tree is given
  by the rate distortion function. The original message is biased with bias $p=0.2$. The rate
  distortion function is always above the AT line and
  thus, the RS solution is always stable.}
  \label{fig:ParityK3P02}
  \vspace{0mm}
\end{figure}

Figure \ref{fig:ParityK3P02} shows the rate-distortion function plotted with the AT stability
line for biased messages with $p=0.2$. All the region below the AT line is unstable.
Since no part of the rate distortion function is under the AT line, the RS solution is
always stable. We did the same experiment for higher values of $K$ and never found
any unstable part.

The lossy compression scheme using a parity tree with non-monotonic hidden units presents
good properties. It saturates the Shannon bound for any value of $K\geq 2$
and the RS solution seems to be always AT stable.

\subsection{AT stability for the committee tree with non-monotonic hidden units}

In the case of a comittee tree with non-monotonic hidden units, we find
\begin{eqnarray}
P & = & R^{-1} \left\{ p \left[ \frac {(e^{\beta}-1) (B_k-B^{*}_k)}
{1+(e^{\beta}-1)B_k} \right]^2 \right.
\nonumber \\
&& + (1-p) \left. \left[ \frac {(e^{\beta}-1) (B^{*}_k-B_k)}
{1+(e^{\beta}-1)(1-B_k)} \right]^2 \right\},
\nonumber \\
\nonumber \\
Q &= & R=P^{\prime} = Q^{\prime} = R^{\prime}=0,
\end{eqnarray}
where
\begin{eqnarray}
B^{*}_k & = & \sum_{\tau_l = \pm 1} \left\{ \Theta \left[ \sum_{l=1}^K \tau_l \right]
\left( \frac {1 + \tau_1 (1 - \frac {4k e^{-k^2/2}} {\sqrt{2 \pi}} - 4 H [ k ])} {2} \right)
\right.
\nonumber \\
&& \left. \times \prod_{l=2}^K \left[ \frac {1 + \tau_l (1 - 4 H [ k ])} {2} \right] \right\}.
\end{eqnarray}
Therefore, using equation (\ref{ATcrit}), the RS stability criterion is given by
\begin{eqnarray}
R & > & K \left\{ p \left[ \frac {(e^{\beta}-1) (B_{\hat{k}}-B^{*}_{\hat{k}})}
{1+(e^{\beta}-1) B_{\hat{k}}} \right]^2 \right.
\nonumber \\
&& + (1-p) \left. \left[ \frac {(e^{\beta}-1) (B^{*}_{\hat{k}}-B_{\hat{k}})}
{1+(e^{\beta}-1) (1-B_{\hat{k}})} \right]^2 \right\},
\end{eqnarray}
where $\beta$ is given by (\ref{betacommitteehidden}), and where $\hat{k}$ satisfies equation
(\ref{khatcommitteehidden}).
\begin{figure}[ht!]
  \vspace{0mm}
  \begin{center}
  \includegraphics[width=0.8\linewidth,keepaspectratio]{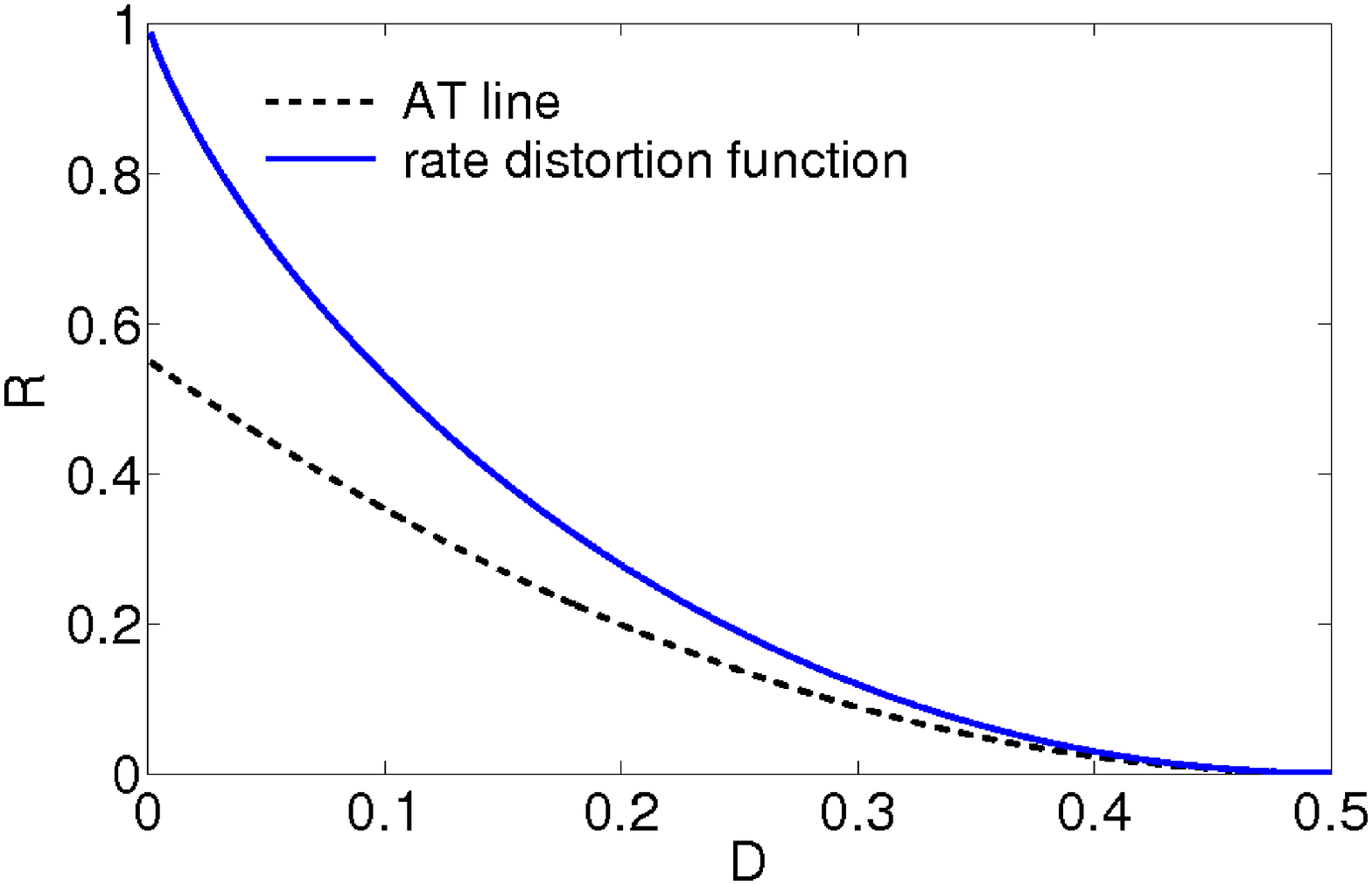}
  \end{center}
  \caption{AT line and rate distortion function for the committee tree with 3 non-monotonic
  hidden units.
  The rate distortion performance of the committee tree is given
  by the rate distortion function.
  The original message is unbiased ($p=0.5$). The rate distortion function is always above
  the AT line and thus, the RS solution is always stable.}
  \label{fig:CommitteeHK3P05}
  \vspace{0mm}
\end{figure}
\begin{figure}[ht!]
  \vspace{0mm}
  \begin{center}
  \includegraphics[width=0.8\linewidth,keepaspectratio]{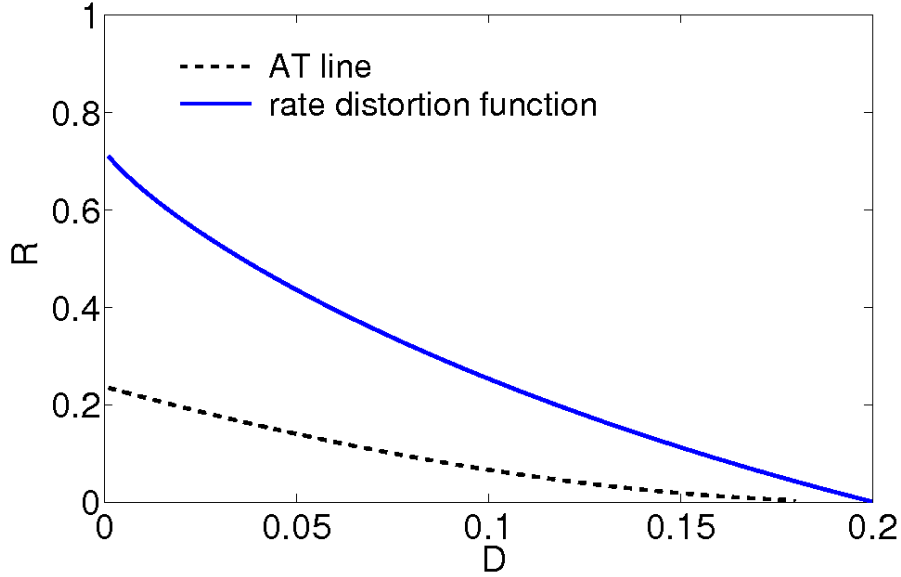}
  \end{center}
  \caption{AT line and rate distortion function for the committee tree with 3 non-monotonic
  hidden units.
  The rate distortion performance of the committee tree is given
  by the rate distortion function.
  The original message is biased ($p=0.2$). The rate distortion function is always above
  the AT line and thus, the RS solution is always stable.}
  \label{fig:CommitteeHK3P02}
  \vspace{0mm}
\end{figure}

However as mentioned in the previous section, it is not clear if there exists $\hat{k}$
which makes equation (\ref{khatcommitteehidden}) true. Nevertheless, we did some numerical
calculations for $K=3$ and $K=5$ (in this case we consider only odd values of $K$ as mentioned
earlier), and always found an optimal $k$ ($\equiv \hat{k})$
in those cases.

We presents here the results obtain for $K=3$.
Figures \ref{fig:CommitteeHK3P05} and \ref{fig:CommitteeHK3P02} show the rate-distortion
function plotted with the AT stability line for unbiased ($p=0.5$) and biased ($p=0.2$)
messages. Since no part of the rate distortion function
is under the AT line, the RS solution is
always stable. We tried also for higher values of $K$ and no unstable part were found
for the RS solution.

The lossy compression scheme using a committee tree with non-monotonic hidden units also
presents good properties. If an optimal $\hat{k}$ exists (which seems to be always true), it
saturates the Shannon bound and the RS solution seems to be always AT stable.

\subsection{AT stability for the committee tree with a non-monotonic output unit}

In the case of a committee tree with a non-monotonic output unit, we find
\begin{eqnarray}
P^{\prime} & = & \frac {R^{-1}} {\pi^2} (1-e^{\beta})^2 2^{-2(K-2)}
\nonumber \\
&& \times \left< \left[ \frac
{\binom{K-2}{\lceil\frac {K-y \sqrt{K-2} k} {2} - 1 \rceil }
- \binom{K-2}{\lfloor\frac {K+y \sqrt{K-2} k} {2} \rfloor }}
{e^{-\beta /2 (y-1)} - y (1-e^{\beta}) C_k }
\right]^2 \right>_y,
\nonumber \\
\nonumber \\
P & = & Q =  R = Q^{\prime} = R^{\prime}=0,
\end{eqnarray}
where $\lceil x \rceil$ denotes the ceiling function
($\lceil x \rceil = \min \{ n \in \mathbb{Z} | n \leq x \}$) and
$\lfloor x \rfloor$ denotes the floor function
($\lfloor x \rfloor = \max \{ n \in \mathbb{Z} | n \leq x \}$).
Therefore, using equation (\ref{ATcrit}), the RS stability criterion is given by
\begin{eqnarray}
R & > & \frac {K(K-1)} {\pi^2} (1-e^{\beta})^2 2^{-2(K-2)}
\nonumber \\
&& \times \left< \left[ \frac
{\binom{K-2}{\lceil\frac {K-y \sqrt{K-2} \hat{k}} {2} - 1 \rceil }
- \binom{K-2}{\lfloor\frac {K+y \sqrt{K-2} \hat{k}} {2} \rfloor }}
{e^{-\beta /2 (y-1)} - y (1-e^{\beta}) C_{\hat{k}} }
\right]^2 \right>_y, \label{ATcommitteeoutput}
\end{eqnarray}
where $\beta$ is given by (\ref{betacommitteeoutput}), and where $\hat{k}$ satisfies equation
(\ref{khatcommitteeoutput}). $<\ldots>_y$ denotes the same expectation as in the parity
tree case.

\begin{figure}[ht!]
  \vspace{0mm}
  \begin{center}
  \includegraphics[width=0.8\linewidth,keepaspectratio]{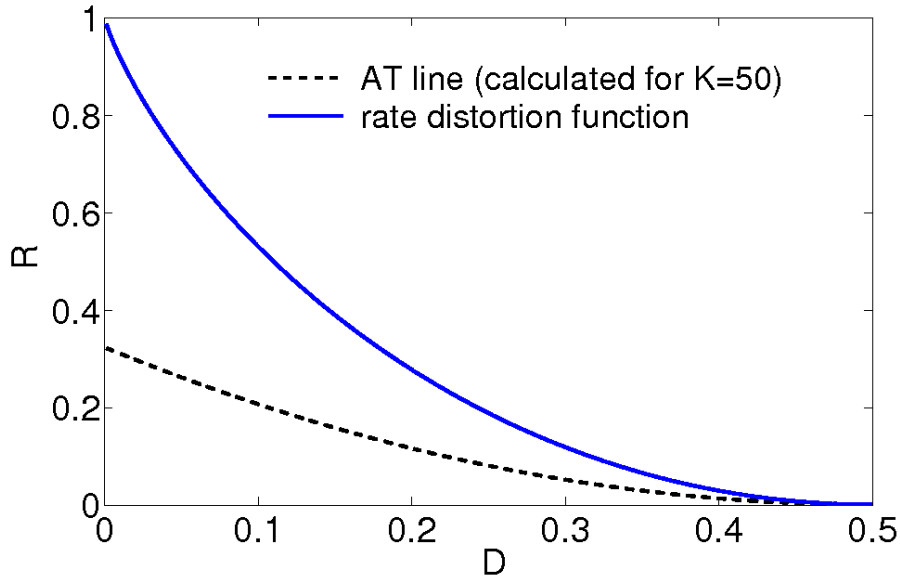}
  \end{center}
  \caption{AT line and rate distortion function for the committee tree with a non-monotonic
  output unit.
  The rate distortion performance of the committee tree is given
  by the rate distortion function.
  The original message is unbiased ($p=0.5$). The AT line
  is calculated using $K=50$. The rate distortion function is always above
  the AT line and thus, the RS solution is always stable.}
  \label{fig:CommitteeOK50P05}
  \vspace{0mm}
\end{figure}

\begin{figure}[ht!]
  \vspace{0mm}
  \begin{center}
  \includegraphics[width=0.8\linewidth,keepaspectratio]{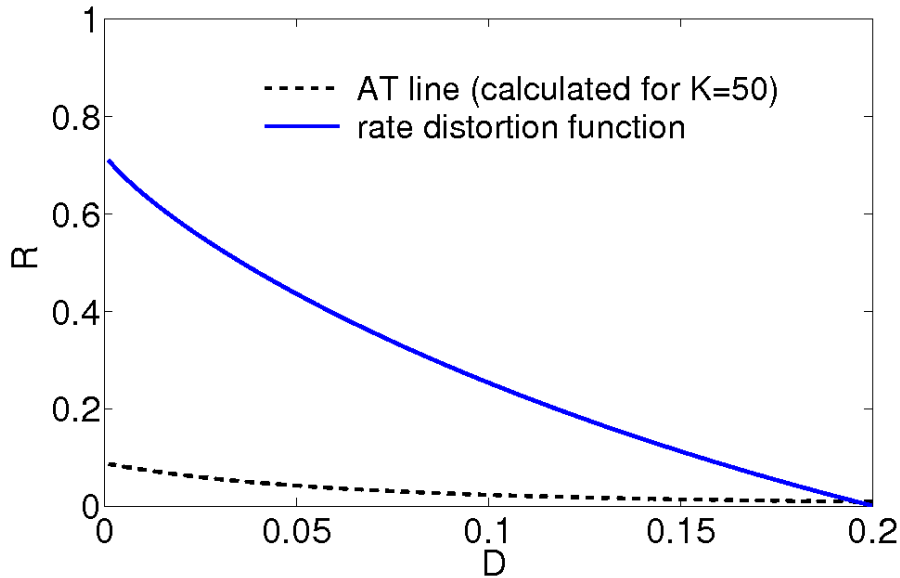}
  \end{center}
  \caption{AT line and rate distortion function for the committee tree with a non-monotonic
  output unit.
  The rate distortion performance of the committee tree is given
  by the rate distortion function.
  The original message is biased ($p=0.2$). The AT line
  is calculated using $K=50$.
  The rate distortion function is always above
  the AT line except for a very narrow region where $D \approx p$. }
  \label{fig:CommitteeOK50P02}
  \vspace{0mm}
\end{figure}

However as mentioned in the previous section, here also it is not clear if there exists
$\hat{k}$ such that equation (\ref{khatcommitteeoutput}) is satisfied. On top of that,
the function $C_k$ which depends on $k$ is not continuous but discrete. $C_k$ is a step
function of $k$. Therefore,
we might have no $k$ satisfying equation (\ref{khatcommitteeoutput}).
On the other hand, since $C_k$ is a step function of $k$, if we find a $k$ which satisfy
equation (\ref{khatcommitteeoutput}),
then it implies that $\hat{k}$ is not given by a unique solution but by
 an optimal interval where
all the elements in this interval satisfy (\ref{khatcommitteeoutput}).
We did some numerical experiments for unbiased message ($p=0.5$). For the special
case of $K=2$, equation (\ref{khatcommitteeoutput}) is clearly satisfied for any
$k \in ] 0, \sqrt{2} [$ so that in this case $\hat{k}$ is given by any element of
the interval $] 0, \sqrt{2} [$. But for
$K >2 $ (we checked until $K=100$), we did not found any optimal $k$. We did the same
thing for biased message ($p=0.2$) with a fixed distortion $D=0.1$ and
for any $K \leq 100$ no optimal $k$ exists.
This implies that in the general case, the committee tree with a non-monotonic
output unit does not saturate the Shannon bound. However, if
the number of hidden units $K$ becomes very large, we can apply the central limit
theorem to replace the scalar product $\mbi{s}_l \cdot \mbi{x}_l$ by a Gaussian variable.
Under these conditions, the expression of $C_k$ becomes very simple,
\begin{equation}
C_k = 1 - 2 H(k).
\end{equation}
In this case, $C_k$ is no more a step function, but a continuous function of $k$
and it is easy to see that there is always a $k$ which satisfy
the equation
\begin{equation}
1 - 2 H(k)= \frac {p-D} {1-2D}. \label{khatCLT}
\end{equation}
Let us denote it by $\hat{k}_{\text{inf}}$.
So in the large $K$ limit, $\hat{k}=\hat{k}_{\text{inf}}$
exists and is unique.
The compression scheme with a committee tree using a non-monotonic output unit
saturates the Shanon bound in this limit.
It is however hard to check the AT stability for an infinite number of hidden units $K$
(the binomial coefficient follows a factorial growth)
but we claim the solution obtained by the RS ansatz to be always AT stable (except
for some very narrow region where $D \approx p$).
We show in Figures \ref{fig:CommitteeOK50P05} and \ref{fig:CommitteeOK50P02}
the rate distortion function plotted with the AT line for
$K=50$ hidden units for unbiased ($p=0.5$) and biased ($p=0.2$) messages.

To sum up this subsection, the lossy compression scheme using a committee tree with
a non-monotonic output unit present a quite complex structure which does not saturate the
Shannon bound in most cases. However, it does saturate it when the number of hidden units
becomes infinite. Concerning the AT stability, the committee tree with
a non-monotonic output unit does not seem to exhibit any critical
instability for the RS solution.

\FloatBarrier

\section{Conclusion and Discussion}

We investigated a lossy compression scheme for uniformly biased
Boolean messages using non-monotonic parity tree and non-monotonic committee tree
multilayer perceptrons. All the schemes were shown to saturate the
Shannon bound under some specific conditions. The replica symmetric
solution is always stable which tends to confirm the validity of the
replica symmetric ansatz.

The Edwards-Anderson order parameter $q$ was always found to be $0$, meaning that codewords are uncorrelated in the codeword space. As already mentioned in \cite{Hosaka2002,Mimura2006}, one may conjecture that this is a necessary condition for a lossy compression scheme to achieve Shannon limit. The mirror symmetry seems then to be an essential feature to saturate the Shannon bound. The committee tree with non-monotonic hidden units corresponds to the same committee tree model as in Mimura et al. paper \cite{Mimura2006} with the exception of the hidden layer transfer function which is given by the non-monotonic transfer function $f_k$ in this paper. By enforcing mirror symmetry in the hidden layer, we were able to get Shannon optimal performance for an infinite length codeword independently of the number of hidden units whereas this was not possible using Mimura et al. model, even for an infinite number of hidden units. In the same way, keeping the monotonic $\sgn$ function as the transfer function of the hidden layer and transforming only the output unit into a non-monotonic one by the use of $f_k$ (i.e.: the committee tree with a non monotonic output unit), we were able to reach Shannon limit with an infinite number of hidden units. Once again, enforcing mirror symmetry enabled to get Shannon optimal performance.

Next, one can easily derive a lower bound for the number of optimal codewords (here optimal means a codeword which gives the minimum achievable distortion of the corresponding scheme) for each of the three schemes. In the case of the parity tree and committee tree with non-monotonic hidden units, there are at least $2^K$ optimal codewords in the codeword space. Indeed, if $\mbi{s}$ denotes an optimal codeword, we can replace any of its component $\mbi{s}_l$ by $-\mbi{s}_l$ without altering the output of the hidden layer and thus leave the output of the network unchanged.
In the case of the committee tree with a non-monotonic output unit, because of the more complex structure of the hidden layer, we can only guaranty the existence of $2$ optimal codewords which are given by $\mbi{s}$ and $-\mbi{s}$.

However, a formal encoder for those schemes would require a computational cost which
grows exponentially with the original message length to perform its task.
We need more efficient algorithms to reduce the encoding time.
A preliminary study made by Hosaka et al. \cite{Hosaka2006}
uses the BP algorithm for this task.
This could be a good solution to achieve the encoding phase in
a polynomial time. Another possibility is to use the survey propagation algorithm approach which was developed for satisfiability problems \cite{Mezard2002}.
Furthermore, as mentioned above, the parity tree and committee tree with non-monotonic hidden units have their number of optimal codeword which grows exponentially with the number of hidden units. This could made the search for one optimal codeword easier to achieve using some proper heuristics.
This issue will be studied
in a next paper.

\vspace*{5mm}
\begin{center}
{\small \bf ACKNOWLEDGEMENTS}
\end{center}

This work was partially supported by a Grant-in-Aid for Encouragement of
Young Scientists (B) (Grant No. 18700230), Grant-in-Aid for Scientific
Research on Priority Areas (Grant Nos. 18079003, 18020007),
Grant-in-Aid for Scientific Research (C) (Grant No. 16500093), and a
Grant-in-Aid for JSPS Fellows (Grant No. 06J06774) from the Ministry of
Education, Culture, Sports, Science and Technology of Japan. T. O. was
partially supported by research fellowship from
the Japan Society for the Promotion of Science.

\vspace*{5mm}
\appendix

\section{Analytical Evaluation using the replica method}
\label{appendix.ReplicaMethod}

The free energy can be evaluated by the replica method (the parameter $k$
is fixed here),
\begin{equation}
f(\beta,R)= - \frac 1 {\beta N} \lim\limits_{n \to 0}
\frac {\left< Z(\beta,\mbi{y},\mbi{x})^n \right>_{\mbi{y},\mbi{x}} - 1} {n}
\end{equation}
where $Z(\beta,\mbi{y},\mbi{x})^n$ denotes the $n$-times replicated partition function
\begin{equation}
Z(\beta,\mbi{y},\mbi{x})^n = \sum_{\mbi{s}^1,\ldots,\mbi{s}^n} \prod_{a =1}^n
\exp \left[ -\beta \mathcal{H} (\mbi{y},\mbi{\hat{y}}(\mbi{s}^{a})) \right]. \label{replicaZ}
\end{equation}
The vector $\mbi{s}^{a}$ is given by
$\mbi{s}^{a}=(\mbi{s}^{a}_1, \ldots ,\mbi{s}^{a}_K)$ and the superscript
$a$ denotes the replica index.

By using the zero entropy criterion \cite{Krauth1989}, we have
\begin{eqnarray}
0 & = & \beta [U - F]
\nonumber \\
u & = & f,
\end{eqnarray}
where $U$ denotes the internal energy, and $F$ the free energy.
$u$ and $f$ denotes the same quantity per bit ($u=U/N, f=F/N$). In the zero
entropy limit, only one state
of the dynamical variable $\mbi{s}$ achieves a distortion per bit inferior or equal to $D$.
The free energy per bit
\begin{equation}
f(\beta,R)=- \frac {1} {\beta N} \ln \left< Z(\beta,\mbi{y},\mbi{x}) \right>_{\mbi{y},\mbi{x}}
\end{equation}
is equal to the internal energy per bit
\begin{equation}
u(\beta) = \frac {\partial \beta f} {\partial \beta}.
\end{equation}
This result $f(\beta,R)=u(\beta)$ gives us an explicit relation
between the code rate $R$ and the inverse temperature $\beta$.

Since this temperature was artificially introduced by means of the parameter $\beta$,
we should get rid of it by taking the zero temperature limit
($\beta \to + \infty$) where the dynamical variable freezes.
At this limit, one can retrieve the codeword which minimizes the free energy and
gives the best achievable code rate.
However, since a distortion per bit $D$ is tolerated, at the zero temperature limit
the internal energy per bit should be equal to this distortion. This motivates the
introduction of the following identity
\begin{equation}
\lim\limits_{\beta \to + \infty} u(\beta) = D. \label{Uident}
\end{equation}
Finally, at this zero temperature limit,
one can get an explicit relation which binds the best achievable
code rate $R$ with the distortion $D$:
\begin{equation}
f(D,R)=D.
\end{equation}

We now proceed to the calculation of the replicated partition function (\ref{replicaZ}).
Inserting the identity
\begin{eqnarray}
1 & = & \prod_{a<b}^{n} \prod_{l=1}^{K} \int_{-\infty}^{+\infty} d q^{ab}_l
\delta \left( \mbi{s}^a_l \cdot \mbi{s}^b_l - \frac N K q^{ab}_l \right)
\nonumber \\
& = & \left( \frac 1 {2 \pi i} \right)^{n(n-1)K/2}
\int \left( \prod_{a<b} \prod_{l} d q^{ab}_l d \hat{q}^{ab}_l  \right)
\nonumber \\
&& \times \exp \left[ \sum_{a<b} \sum_{l} \hat{q}^{ab}_l
 \left( \mbi{s}^a_l \cdot \mbi{s}^b_l - \frac N K q^{ab}_l \right) \right]
\end{eqnarray}
into (\ref{replicaZ}) enables to separate the relevant order parameter, and
to calculate the average moment
$\left< Z(\beta,\mbi{y},\mbi{x})^n \right>_{\mbi{y},\mbi{x}}$ for natural numbers
n as,
\begin{widetext}
\begin{eqnarray}
\left< Z(\beta,\mbi{y},\mbi{x})^n \right>_{\mbi{y},\mbi{x}} & \backsimeq &
\int \left( \prod_{a<b} \prod_{l} d q^{ab}_l d \hat{q}^{ab}_l  \right)
\nonumber \\
&& \times \exp N \left[ R^{-1} \ln \left<
\int \left( \prod_{l} d \mbi{u}_l d \mbi{v}_l \right)
\prod_{l} \{ e^{-(1/2) {}^t \mbi{v}_l \mathcal{Q}_l \mbi{v}_l + i \mbi{v}_l \cdot \mbi{u}_l} \}
\right. \right. \nonumber \\
&& \times
\left. \prod_{a} \{ e^{- \beta} + (1 - e^{-\beta}) \Theta (y,\{ u^a_l \} ) \} \right>_y
\nonumber \\
&& \left. + \frac 1 K \ln \sum_{\{ s^a_l \} } \exp \left( \sum_{a<b} \sum_{l}
\hat{q}^{ab}_l s^a_l s^b_l \right) - \frac 1 K \sum_{a<b} \sum_{l}
q^{ab}_l \hat{q}^{ab}_l
\right] , \label{zgeneral}
\end{eqnarray}
\end{widetext}
where $\mathcal{Q}_l$ is a $n \times n$ matrix having elements $\{ q^{ab}_l \}$
and where $< \ldots >_y$ denotes the expectation with respect to (\ref{ydistrib}).
The function $\Theta (y,\{ u^a_l \} )$ depends on the decoder and will be
discussed in the following subsections. We analyze the scheme in the thermodynamic
limit $N,M \to + \infty$, while the code rate $R$ is kept finite. In this
limit, (\ref{zgeneral}) can be evaluated using the saddle point method with respect
to $q^{ab}_l$ and $\hat{q}^{ab}_l$. To continue the calculation, we have to make some
assumptions about the structure of these order parameters.
We use here the so-called replica symmetric (RS) ansatz
\begin{eqnarray}
q^{ab}_l & = & (1-q) \delta_{ab} + q,
\nonumber \\
\hat{q}^{ab}_l & = & (1- \hat{q} ) \delta_{ab} + \hat{q},
\end{eqnarray}
where $\delta_{ab}$ denotes the Kronecker delta. This ansatz means that all the hidden
units are equivalent after averaging over the disorder.

\subsection{Replica symmetric evaluation for the parity tree with non-monotonic hidden units}

In the case of a parity tree with non-monotonic hidden units, the function
$\Theta (y,\{ u^a_l \} )$ in (\ref{zgeneral}) is given by
\begin{eqnarray}
\Theta (y,\{ u^a_l \} ) = \theta \left( y \prod_{l=1}^{K} \sgn
\left[ k^2 - (u^a_l)^2 \right] \right) .
\end{eqnarray}
We can then obtain the expression of the free energy as
\begin{eqnarray}
f(\beta,R,k,q,\hat{q}) & = & - \frac 1 {\beta} \underset{q,\hat{q}}{\text{extr}}
\left\{ \left< \int_{-\infty}^{+\infty} \left( \prod_{l=1}^K D t_l \right)
\right. \right. \nonumber \\ &&
\times \ln \left[ e^{-\beta} + (1 - e^{-\beta}) \Pi_k ( \{ t_l \} ,y) \right] \Bigg>_y
\nonumber \\ &&
+ R \int_{-\infty}^{+\infty} D u \ln \left[ 2 \cosh (\sqrt{\hat{q}} u) \right]
\nonumber \\ &&
- R \frac {\hat{q} (1-q)} {2}
\Bigg\} , \label{parityenergy}
\end{eqnarray}
where
\begin{eqnarray}
Dx & = & \frac{ e^{-x^2/2} dx} {\sqrt{2 \pi}} ,
\nonumber \\
\Pi_k ( \{ t_l \} ,y) & = & \frac 1 2 + \frac y 2 \prod_{l=1}^K \left\{
1 -2 H \left[ \frac {k+\sqrt{q} t_l} {\sqrt{1-q}} \right]
\right. \nonumber \\ && \makebox[2cm]{} \left.
-2 H \left[ \frac {k-\sqrt{q} t_l} {\sqrt{1-q}} \right]
\right\} .
\end{eqnarray}
Taking the derivative of (\ref{parityenergy}) with respect to $q$ and $\hat{q}$
gives the saddle point equations for the order parameters
\begin{eqnarray}
\hat{q} & = & 2 R^{-1} \left<  \int_{-\infty}^{+\infty} \left( \prod_{l=1}^K D t_l \right)
\right. \nonumber \\ && \makebox[1cm]{} \left. \times
\frac {-(1-e^{-\beta}) \Pi_k^{\prime} ( \{ t_l \} ,y)}
{e^{-\beta}+(1-e^{-\beta}) \Pi_k ( \{ t_l \} ,y)} \right>_y ,
\nonumber \\
q & = & \int_{-\infty}^{+\infty} Du \tanh^2 (\sqrt{\hat{q}}u),
\end{eqnarray}
where $\Pi_k^{\prime} ( \{ t_l \} ,y)=\partial \Pi_k ( \{ t_l \} ,y)/\partial q$.

We solved this saddle point equation numerically and find
that the solution is given for $q=\hat{q}=0$. According
to \cite{Hosaka2002,Mimura2006} this result was expected and implies
that all the codewords are uncorrelated and distributed all around $\mathcal{S}^N$.
Substituting $q=\hat{q}=0$ into (\ref{parityenergy}), one can finally
find the free energy given by (\ref{parityRSenergy}).

\subsection{Replica symmetric evaluation for the committee tree with non-monotonic hidden
units}

In the case of a committee tree with non-monotonic hidden units, the function
$\Theta (y,\{ u^a_l \} )$ in (\ref{zgeneral}) is given by
\begin{eqnarray}
\Theta (y,\{ u^a_l \} ) = \theta \left( y \sum_{l=1}^{K} \sgn
\left[ k^2 - (u^a_l)^2 \right] \right) .
\end{eqnarray}
We can then obtain the expression of the free energy as
\begin{eqnarray}
f(\beta,R,k,q,\hat{q}) & = & - \frac 1 {\beta} \underset{q,\hat{q}}{\text{extr}}
\left\{ \left< \int_{-\infty}^{+\infty} \left( \prod_{l=1}^K D t_l \right)
\right. \right. \nonumber \\ && \times
\ln \left[ e^{-\beta} + (1 - e^{-\beta}) \Sigma_k ( \{ t_l \} ,y) \right] \Bigg>_y
\nonumber \\ &&
+ R \int_{-\infty}^{+\infty} D u \ln \left[ 2 \cosh (\sqrt{\hat{q}} u) \right]
\nonumber \\ &&
- R \frac {\hat{q} (1-q)} {2}
\Bigg\} , \label{committeeHenergy}
\end{eqnarray}
where
\begin{eqnarray}
\Sigma_k ( \{ t_l \} ,y) & = & \sum_{\tau_l = \pm 1}
\left\{
\theta \left[ y \sum_{l=1}^K \tau_l \right] \right.
\nonumber \\ && \times
\prod_{l=1}^K \left[ \frac {1+\tau_l} {2}
- \tau_l H \left( \frac {k+\sqrt{q} t_l} {\sqrt{1-q}} \right)
\right. \nonumber \\ && \left. \left.
- \tau_l H \left( \frac {k-\sqrt{q} t_l} {\sqrt{1-q}} \right)
\right]
\right\}.
\end{eqnarray}
Taking the derivative of (\ref{committeeHenergy}) with respect to $q$ and $\hat{q}$
gives the saddle point equations for the order parameters
\begin{eqnarray}
\hat{q} & = & 2 R^{-1} \left<  \int_{-\infty}^{+\infty} \left( \prod_{l=1}^K D t_l \right)
\right. \nonumber \\ && \makebox[2cm]{} \left. \times
\frac {-(1-e^{-\beta}) \Sigma_k^{\prime} ( \{ t_l \} ,y)}
{e^{-\beta}+(1-e^{-\beta}) \Sigma_k ( \{ t_l \} ,y)} \right>_y ,
\nonumber \\
q & = & \int_{-\infty}^{+\infty} Du \tanh^2 (\sqrt{\hat{q}}u),
\end{eqnarray}
where $\Sigma_k^{\prime} ( \{ t_l \} ,y)=\partial \Sigma_k ( \{ t_l \} ,y)/\partial q$.

We solved this saddle point equation numerically and here also we find
that the solution is given for $q=\hat{q}=0$.
Substituting $q=\hat{q}=0$ into (\ref{committeeHenergy}), one can finally
find the free energy given by (\ref{committeeHRSenergy}).

\subsection{Replica symmetric evaluation for the committee tree with a non-monotonic output
unit}

In the case of a committee tree with a non-monotonic output unit, the function
$\Theta (y,\{ u^a_l \} )$ in (\ref{zgeneral}) is given by
\begin{eqnarray}
\Theta (y,\{ u^a_l \} ) = \theta \left( y  \left[ k^2
- \frac 1 K \left( \sum_{l=1}^{K} \sgn [u^a_l] \right)^2 \right] \right) .
\end{eqnarray}
We can then obtain the expression of the free energy as
\begin{eqnarray}
f(\beta,R,k,q,\hat{q}) & = & - \frac 1 {\beta} \underset{q,\hat{q}}{\text{extr}}
\left\{ \left< \int_{-\infty}^{+\infty} \left( \prod_{l=1}^K D t_l \right)
\right. \right. \nonumber \\ && \times
\ln \left[ e^{-\beta} + (1 - e^{-\beta}) F_{\Sigma,k} ( \{ t_l \} ,y) \right] \Bigg>_y
\nonumber \\ &&
+ R \int_{-\infty}^{+\infty} D u \ln \left[ 2 \cosh (\sqrt{\hat{q}} u) \right]
\nonumber \\ &&
- R \frac {\hat{q} (1-q)} {2}
\Bigg\} , \label{committeeOenergy}
\end{eqnarray}
where
\begin{eqnarray}
F_{\Sigma,k} ( \{ t_l \} ,y) & = & \sum_{\tau_l = \pm 1}
\left\{
\theta \left[ y k^2 - \frac y K \Big( \sum_{l} \tau_l \Big)^2 \right] \right.
\nonumber \\ && \left. \times
\prod_{l=1}^K  H \left[ - t_l \tau_l \sqrt{\frac {q} {1-q}} \right]
\right\} .
\end{eqnarray}
Taking the derivative of (\ref{committeeOenergy}) with respect to $q$ and $\hat{q}$
gives the saddle point equations for the order parameters
\begin{eqnarray}
\hat{q} & = & 2 R^{-1} \left<  \int_{-\infty}^{+\infty} \left( \prod_{l=1}^K D t_l \right)
\right. \nonumber \\ && \makebox[2cm]{} \left. \times
\frac {-(1-e^{-\beta}) F_{\Sigma,k}^{\prime} ( \{ t_l \} ,y)}
{e^{-\beta}+(1-e^{-\beta}) F_{\Sigma,k} ( \{ t_l \} ,y)} \right>_y ,
\nonumber \\
q & = & \int_{-\infty}^{+\infty} Du \tanh^2 (\sqrt{\hat{q}}u),
\end{eqnarray}
where
$F_{\Sigma,k}^{\prime} ( \{ t_l \} ,y)=\partial F_{\Sigma,k} ( \{ t_l \} ,y)/\partial q$.

We solved this saddle point equation numerically and here also we find
that the solution is given for $q=\hat{q}=0$.
Substituting $q=\hat{q}=0$ into (\ref{committeeOenergy}), one can finally
find the free energy given by (\ref{committeeORSenergy}).

\section{Almeida-Thouless stability criterion}
\label{appendix.AT}

The Hessian computed at the RS saddle point characterizes fluctuations in
the order parameters $q^{ab}_l$ and $\hat{q}^{ab}_l$ around the RS saddle point. Instability
of the RS solution is signaled by a change of sign of at least one of the eigenvalues
of the Hessian. Let $\mathcal{M}( \{ q^{ab}_l \} , \{ \hat{q}^{ab}_l \} )$ be the exponent of
the integrand of integral (\ref{zgeneral}). Equation (\ref{zgeneral}) can be represented
as
\begin{eqnarray}
\left< Z(\beta,\mbi{y},\mbi{x})^n \right>_{\mbi{y},\mbi{x}} & = &
\int \left( \prod_{a<b} \prod_{l} d q^{ab}_l d \hat{q}^{ab}_l  \right)
\nonumber \\
&& \times \exp \left( N \mathcal{M}( \{ q^{ab}_l \} , \{ \hat{q}^{ab}_l \} ) \right) .
\end{eqnarray}
We expand $\mathcal{M}$ around $q$ and $\hat{q}$
in $\delta q^{ab}_l$ and $\delta \hat{q}^{ab}_l$ and then find up the second order
\begin{eqnarray}
\mathcal{M}( \{ q + \delta q^{ab}_l \} , \{ \hat{q} + \delta \hat{q}^{ab}_l \} )
& = & \mathcal{M}( \{ q \} , \{ \hat{q} \} ) + \frac 1 2 {}^t \mbi{v} \mbi{G} \mbi{v} \nonumber
\\
&& + \mathcal{O}
(\| \mbi{v} \|^3),
\end{eqnarray}
where
\begin{equation}
\mbi{v} = {}^t ( \{ \delta q^{ab}_1 \} , \{ \delta \hat{q}^{ab}_1 \} , \ldots , \{ \delta q^{ab}_K \} , \{ \delta \hat{q}^{ab}_K \} )
\end{equation}
is the perturbation to the RS saddle point. The Hessian $\mbi{G}$ is the following
$[ K n(n-1) ] \times [ K n(n-1) ]$ matrix,
\begin{gather}
\mbi{G} =
\begin{pmatrix}
\mbi{U} & \mbi{V} & \ldots & \mbi{V}
\\
\mbi{V} & \mbi{U} & \ldots & \mbi{V}
\\
\vdots  & \vdots  & \ddots & \vdots
\\
\mbi{V} & \mbi{V} & \ldots & \mbi{U}
\end{pmatrix},
\end{gather}
where $n(n-1) \times n(n-1)$ matrices $\mbi{U}$ and $\mbi{V}$ are
\begin{gather}
\mbi{U} =
\begin{pmatrix}
\{ \mbi{U}^{ab,cd} \} & \{ \mbi{\tilde{U}}^{ab,cd} \}
\\
\{ \mbi{\tilde{U}}^{ab,cd} \} &  \{ \mbi{\hat{U}}^{ab,cd} \}
\end{pmatrix},
\\
\\
\mbi{V} =
\begin{pmatrix}
\{ \mbi{V}^{ab,cd} \} & \{ \mbi{\tilde{V}}^{ab,cd} \}
\\
\{ \mbi{\tilde{V}}^{ab,cd} \} &  \{ \mbi{\hat{V}}^{ab,cd} \}
\end{pmatrix},
\end{gather}
with
\begin{equation}
\begin{array}{c}
\mbi{U}^{ab,cd} =
\partial^2 \mathcal{M} / \partial q^{ab}_l \partial q^{cd}_l,
\\
\\
\mbi{\hat{U}}^{ab,cd} =
\partial^2 \mathcal{M} / \partial \hat{q}^{ab}_l \partial \hat{q}^{cd}_l,
\\
\\
\mbi{\tilde{U}}^{ab,cd} =
\partial^2 \mathcal{M} / \partial q^{ab}_l \partial \hat{q}^{cd}_l,
\\
\\
\mbi{V}^{ab,cd} =
\partial^2 \mathcal{M} / \partial q^{ab}_l
\partial q^{cd}_{l^{\prime}} \quad (l \neq l^{\prime}),
\\
\\
\mbi{\hat{V}}^{ab,cd} =
\partial^2 \mathcal{M} / \partial \hat{q}^{ab}_l \partial
\hat{q}^{cd}_{l^{\prime}} \quad (l \neq l^{\prime}),
\\
\\
\mbi{\tilde{V}}^{ab,cd} =
\partial^2 \mathcal{M} / \partial q^{ab}_l \partial
\hat{q}^{cd}_{l^{\prime}} \quad (l \neq l^{\prime}) .
\end{array}
\end{equation}
For $q, \hat{q}$ to be a local maximum of $\mathcal{M}$, it is necessary for the Hessian
$\mbi{G}$ to be negative definite (i.e.: all of its eigenvalues must be negative).

To check these eigenvalues, we use the same method as in \cite{Mimura2006}. We do not
give the mathematical details here. Finally, using Gardner's method \cite{Gardner1988},
we can derive the stability criterion for the RS solution to be stable as
\begin{equation}
K \gamma < 1, \label{ATcrit}
\end{equation}
where
\begin{equation}
\begin{array}{c}
\gamma \equiv \gamma_0 + (K-1) \gamma_1,
\\
\\
\gamma_0 \equiv P - 2Q + R,
\\
\\
\gamma_1 \equiv P^{\prime} - 2Q^{\prime} + R^{\prime},
\\
\\
P \equiv \mbi{U}^{ab,ab},
\\
\\
Q \equiv \mbi{U}^{ab,ac} \quad (b \neq c),
\\
\\
R \equiv \mbi{U}^{ab,cd} \quad (a\neq c, b \neq d),
\\
\\
P^{\prime} \equiv \mbi{V}^{ab,ab},
\\
\\
Q^{\prime} \equiv \mbi{V}^{ab,ac} \quad (b \neq c),
\\
\\
R^{\prime} \equiv \mbi{V}^{ab,cd} \quad (a\neq c, b \neq d) .
\end{array}
\end{equation}
The line $K \gamma = 1$ defines the so called AT line.



\begin{thebibliography}{26}

\bibitem{Sourlas1989} N. Sourlas, Nature, {\bf 339}, 693 (1989). 

\bibitem{Kabashima2000} Y. Kabashima, T. Murayama and D. Saad, Phys. Rev. Lett., {\bf 84},
1355 (2000). 

\bibitem{Nishimori1999} H. Nishimori and K. Y. M. Wong, Phys. Rev. E, {\bf 60}, 132 (1999).

\bibitem{Montanari2000} A. Montanari and N. Sourlas, Eur. Phys. J. B, {\bf 18}, 107 (2000). 

\bibitem{Tanaka2001} T. Tanaka, Europhys. Lett., {\bf 54}, 540 (2001). 

\bibitem{Tanaka2005} T. Tanaka and M. Okada, IEEE Trans. Inform. Theory, {\bf 51}, 2, 700 (2005). 

\bibitem{Murayama2003} T. Murayama and M. Okada, J. Phys. A: Math. Gen.,{\bf 36}, 11123 (2003). 

\bibitem{Murayama2004} T. Murayama, Phys. Rev. E, {\bf 69}, 035105(R) (2004). 

\bibitem{Hosaka2002} T. Hosaka, Y. Kabashima and H. Nishimori, Phys. Rev. E, {\bf 66}, 066126
(2002). 

\bibitem{Hosaka2005} T. Hosaka and Y. Kabashima, J. Phys. Soc. Jpn., {\bf 74}, 1, 488 (2005). 

\bibitem{Mimura2006} K. Mimura and M. Okada, Phys. Rev. E, {\bf 74}, 026108 (2006).


\bibitem{Shannon1948} C. E. Shannon, Bell Syst. Tech. J., {\bf 27}, 379 (1948).

\bibitem{Gallager1962} R. Gallager, IRE Trans. on Info. Theory, 1968, \textbf{IT-8}, 21 (1968) 

\bibitem{MacKay1997} D.J.C. MacKay, R.M. Neal, IEEE Electronics Letters, {\bf 33}(6), 457 (1997)

\bibitem{Richardson2001} T.J. Richardson, R.L. Urbanke, IEEE Trans. Inform. Theory, {\bf 47}(2), 638 (2001)

\bibitem{Shannon1959} C. E. Shannon, IRE Nat. Conv. Rec., {\bf 4}, 142 (1959).

\bibitem{Matsunaga2003} Y. Matsunaga, H. Yamamoto, IEEE Trans. Inform. Theory, {\bf 49}(9), 2225 (2003)

\bibitem{Martinian2006} E. Martinian, M.J. Wainwright, Data Compression Conference, (2006)

\bibitem{Wainwright2007} M.J. Wainwright, IEEE Signal Processing Magazine, {\bf 24}(5), 47 (2007)

\bibitem{Hosaka2006} T. Hosaka, Y. Kabashima, Physica A, {\bf 365}, 113, (2006)

\bibitem{Cover1991} T. M. Cover and J. A. Thomas, Elements of Information Theory (Wiley, New York, 1991). 

\bibitem{Gardner1988} E. Gardner, J. Phys. A: Math. Gen., {\bf 21}, 257 (1988). 

\bibitem{Krauth1989} W. Krauth and M. M\'{e}zard, J. Phys. (France), {\bf 50}, 3057 (1989).

\bibitem{Engel2001} A. Engel and C. Van den Broeck, Statistical Mechanics of Learning (Cambridge University Press, 2001). 

\bibitem{Almeida1978} J. R. L. de Almeida and D. J. Thouless, J. Phys. A: Math. Gen., {\bf 11}(5), 983 (1978). 

\bibitem{Mezard2002} M. M\'{e}zard, G. Parisi, R. Zecchina, Science, {\bf 297}, 812 (2002)  

\end{thebibliography}
\end{document}